\documentclass[10pt]{article}

\usepackage{amsmath}
\usepackage{amssymb}

\usepackage{graphicx}

\usepackage{cite}

\usepackage{color} 

\usepackage{booktabs}

\usepackage{ulem}

\usepackage[labelfont=bf,labelsep=period,justification=raggedright]{caption}

\makeatletter
\renewcommand{\@biblabel}[1]{\quad#1.}
\makeatother

\date{}

\begin{document}

\begin{flushleft}
{\Large
\textbf{Ovarian volume throughout life: a validated normative model.}
}
\\
Thomas W. Kelsey$^{1}$, 
Sarah K. Dodwell$^{2}$, 
A. Graham Wilkinson$^{3}$,
Tine Greve$^{4}$,
Claus Y. Andersen$^{4}$,
Richard A. Anderson$^{5}$,
W. Hamish B. Wallace$^{6\ast}$
\\
\bf{1} School of Computer Science, University of St Andrews, St Andrews, Fife, United Kingdom
\\
\bf{2} School of Medicine, University of Edinburgh, Edinburgh, United Kingdom
\\
\bf{3} Department of Paediatric Radiology, Royal Hospital for Sick Children, Edinburgh, United Kingdom
\\
\bf{4} Laboratory of Reproductive Biology, Section 5712, The Juliane Marie Centre for Women, Children and Reproduction, University Hospital of Copenhagen, University of Copenhagen, DK-2100 Copenhagen, Denmark
\\
\bf{5} MRC Centre for Reproductive Health, Queens Medical Research Institute, University of Edinburgh, Edinburgh, United Kingdom
\\
\bf{6} Department of Haematology/Oncology, Royal Hospital for Sick Children, Edinburgh, United Kingdom

\end{flushleft}

\section*{Abstract}

The measurement of ovarian volume has been shown to be a useful indirect indicator of the ovarian reserve in women of reproductive age, in the diagnosis and management of a number of disorders of puberty and adult reproductive function, and is under investigation as a screening tool for ovarian cancer. To date there is no normative model of ovarian volume throughout life. By searching the published literature for ovarian volume in healthy females, and using our own data from multiple sources (combined n = 59,994) we have generated and robustly validated the first model of ovarian volume from conception to 82 years of age. This model shows that 69\% of the variation in ovarian volume is due to age alone. We have shown that in the average case ovarian volume rises from 0.7 mL (95\% CI 0.4 -- 1.1 mL) at 2 years of age to a peak of 7.7 mL  (95\% CI 6.5 -- 9.2 mL) at 20 years of age with a subsequent decline to about 2.8mL (95\% CI 2.7 -- 2.9 mL) at the menopause and smaller volumes thereafter. Our model allows us to generate normal values and ranges for ovarian volume throughout life. This is the first validated normative model of ovarian volume from conception to old age; it will be of use in the diagnosis and management of a number of diverse gynaecological and reproductive conditions in females from birth to menopause and beyond.


\section*{Introduction}

The main functions of the ovary are to provide gametes and sex steroids to allow and support the establishment of pregnancy, and act as a repository for the non-growing follicles (NGFs) that allow this process to take place over several decades. The main constituents of the ovary are therefore its follicle endowment (both growing and non-growing), and the stromal tissues that support these functions. The human ovary establishes its complete complement of non-growing follicles during fetal life, and after birth there is a continuous process of recruitment until menopause at an average age of 50-51 years, when fewer than one thousand remain \cite{Wallace2010,Hansen2008a,Faddy1996}. There is a wide variation in the age at menopause between individuals \cite{Treloar1981,VanNoord1997} and it is thought that this is due in large part to variations in the initial endowment of NGFs \cite{Wallace2010}. Currently, clinical assessment is unable to assess reliably the number of NGFs, or their rate of loss or activation. 

Ovarian volume is one of several putative biomarkers of the ovarian reserve, others include serum anti-M\"{u}llerian Hormone (AMH), and antral follicle count (AFC) which have have been shown to have clinical utility in the assessment of women with subfertility \cite{Broekmans2006}. Ovarian volume is currently one of the diagnostic criteria for the most common endocrinopathy in women (polycystic ovary syndrome; PCOS) \cite{Rotterdam2004,Rotterdam2004a} and may be of value in screening for ovarian cancer \cite{vannagell2007}.  We have shown a strong and positive correlation between ovarian volume and NGF population in the human ovary for ages 25-51 years \cite{Kelsey2012}, but there is only sparse information available on ovarian volume in healthy young girls and women \cite{Salardi1985}. A greater understanding of the changes in ovarian volume throughout life are likely to be helpful in the diagnosis and treatment of many disorders in gynaecology and reproductive medicine \cite{Lass1999}. 

The data on ovarian volume in young girls is limited due to the lack of an easy non-invasive method of imaging the ovaries accurately. Much of the data that is published is in girls with abnormalities in pubertal development and so does not reflect the healthy population \cite{Bridges1993, Stanhope1985}. In the adult woman the advent of transvaginal ultrasound as a routine gynaecological technique has led to a large source of data on ovarian volume in healthy women \cite{Pavlik2000}. To date no single study has examined ovarian volume across the lifespan in healthy females. The aim of this study is to develop a validated model of ovarian volume in healthy females from conception throughout life from data aggregation from multiple sources.

\begin{figure}[!ht]
\begin{center}
\includegraphics[width=\textwidth]{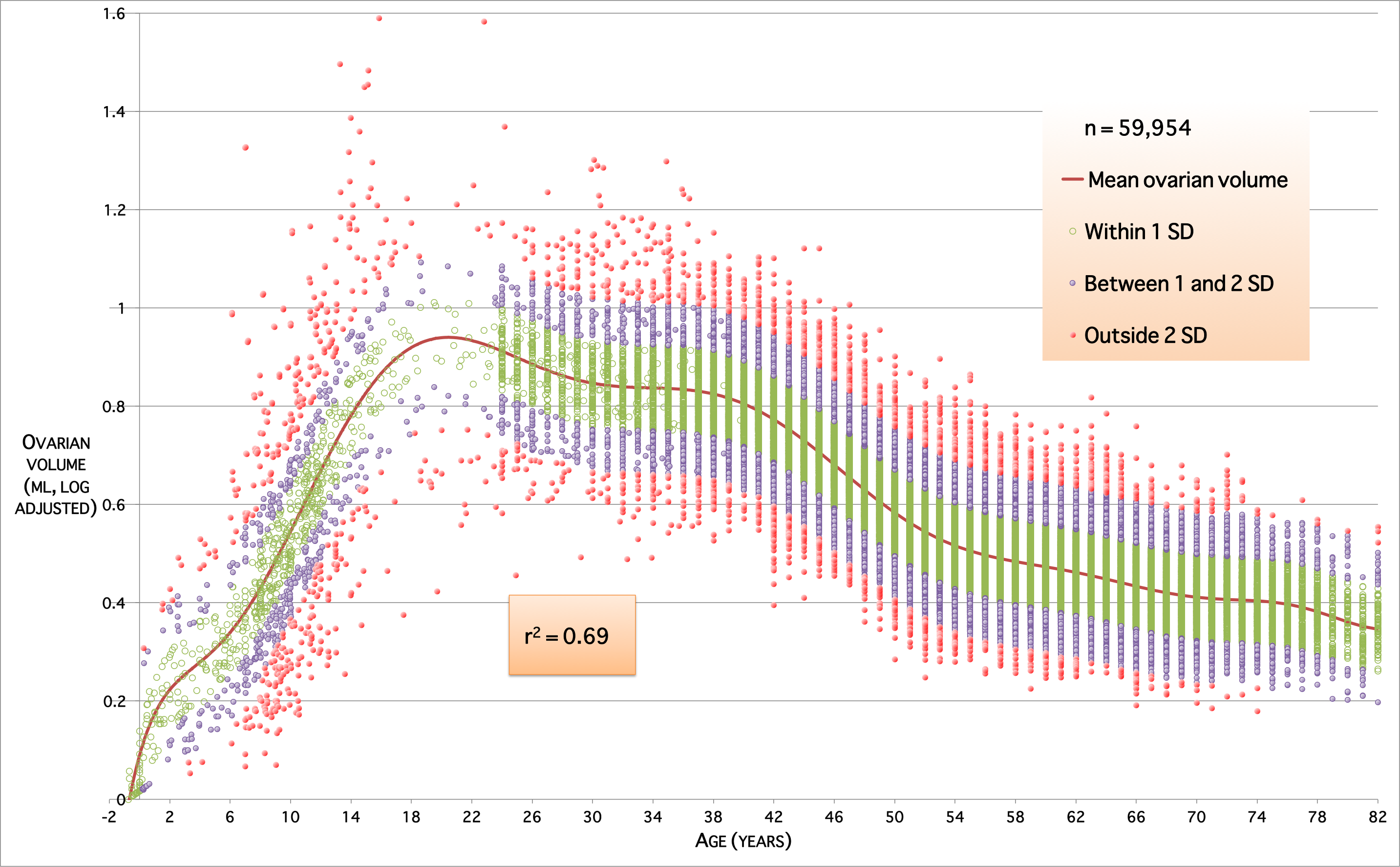}
\end{center}
\caption{
{\bf The validated model of log-adjusted ovarian volume throughout life.} The $r^2$ coefficient of determination indicates that 69\% of the variation in human ovarian volumes is due to age alone. Colour bands indicate ranges within $\pm 1$ standard deviation from mean, within $\pm 1$ and $\pm 2$ standard deviations, and outside $2$ standard deviations.
}
\label{F01}
\end{figure}
\section*{Results}

The validated model is  a degree 14 polynomial of the form
$$
log_{10}(Volume + 1) = c_0 + c_1\mathrm{age}  + c_2\mathrm{age}^2 + \ldots +  c_{13}\mathrm{age}^{14}
$$
with coefficients $c_i$ given in Table \ref{Tcoeff},  and relationship to the data given in Figure \ref{F01}. The model has coefficient of determination $r^2 = 0.69$ indicating that around 69\% of the variation in ovarian volumes throughout life is due to age alone. The residual plot (Figure \ref{F03}) shows a distribution close to the ideal Gaussian curve ($r^2 = 0.993$), this coefficient of determination being higher than that for three other possible curves for these residuals. Moreover, the proportions of residuals within one, two and three standard deviations (respectively 71\%, 96\% and 99\%) are close to the expected values for data with a Gaussian distribution (respectively 68\%, 95\% and 99\%). Figure \ref{F04} is an exemplar of the 5-fold validation process in which a model is chosen that neither overfits nor underfits the underlying dataset.   

  \begin{table}[!ht]
\caption{
\bf{Coefficients for the validated model}}
\begin{tabular}{c|ccccc}
        \hline
 $i$	&	Coefficient $c_i$	&	Standard Error	&	T Value	&	95\% Conf Lim	&	95\% Conf Lim	\\
 \hline
0	&	8.92E-02	&	8.00E-03	&	11.2	&	-1.24E-02	&	1.91E-01	\\
1	&	1.10E-01	&	8.28E-03	&	13.3	&	5.22E-03	&	2.16E-01	\\
2	&	-3.05E-02	&	5.92E-03	&	-5.2	&	-1.06E-01	&	4.47E-02	\\
3	&	5.09E-03	&	1.63E-03	&	3.1	&	-1.56E-02	&	2.58E-02	\\
4	&	-4.35E-04	&	2.34E-04	&	-1.9	&	-3.40E-03	&	2.53E-03	\\
5	&	2.49E-05	&	2.03E-05	&	1.2	&	-2.33E-04	&	2.83E-04	\\
6	&	-1.23E-06	&	1.15E-06	&	-1.1	&	-1.59E-05	&	1.34E-05	\\
7	&	5.43E-08	&	4.49E-08	&	1.2	&	-5.17E-07	&	6.25E-07	\\
8	&	-1.89E-09	&	1.23E-09	&	-1.5	&	-1.75E-08	&	1.38E-08	\\
9	&	4.66E-11	&	2.40E-11	&	1.9	&	-2.59E-10	&	3.52E-10	\\
10	&	-7.87E-13	&	3.31E-13	&	-2.4	&	-4.99E-12	&	3.42E-12	\\
11	&	8.87E-15	&	3.15E-15	&	2.8	&	-3.12E-14	&	4.89E-14	\\
12	&	-6.36E-17	&	1.97E-17	&	-3.2	&	-3.14E-16	&	1.87E-16	\\
13	&	2.63E-19	&	7.32E-20	&	3.6	&	-6.67E-19	&	1.19E-18	\\
14	&	-4.77E-22	&	1.22E-22	&	-3.9	&	-2.02E-21	&	1.07E-21	\\
  \hline
       \end{tabular} 
\begin{flushleft}
 Coefficients for the validated normative model of human ovarian volume throughout life.  Each coefficient value is reported together with estimates of the standard error, T-statistic and 95\% confidence limits for the value.    
\end{flushleft}
\label{Tcoeff}
 \end{table}

\begin{figure}[!ht]
\begin{center}
\includegraphics[width=\textwidth]{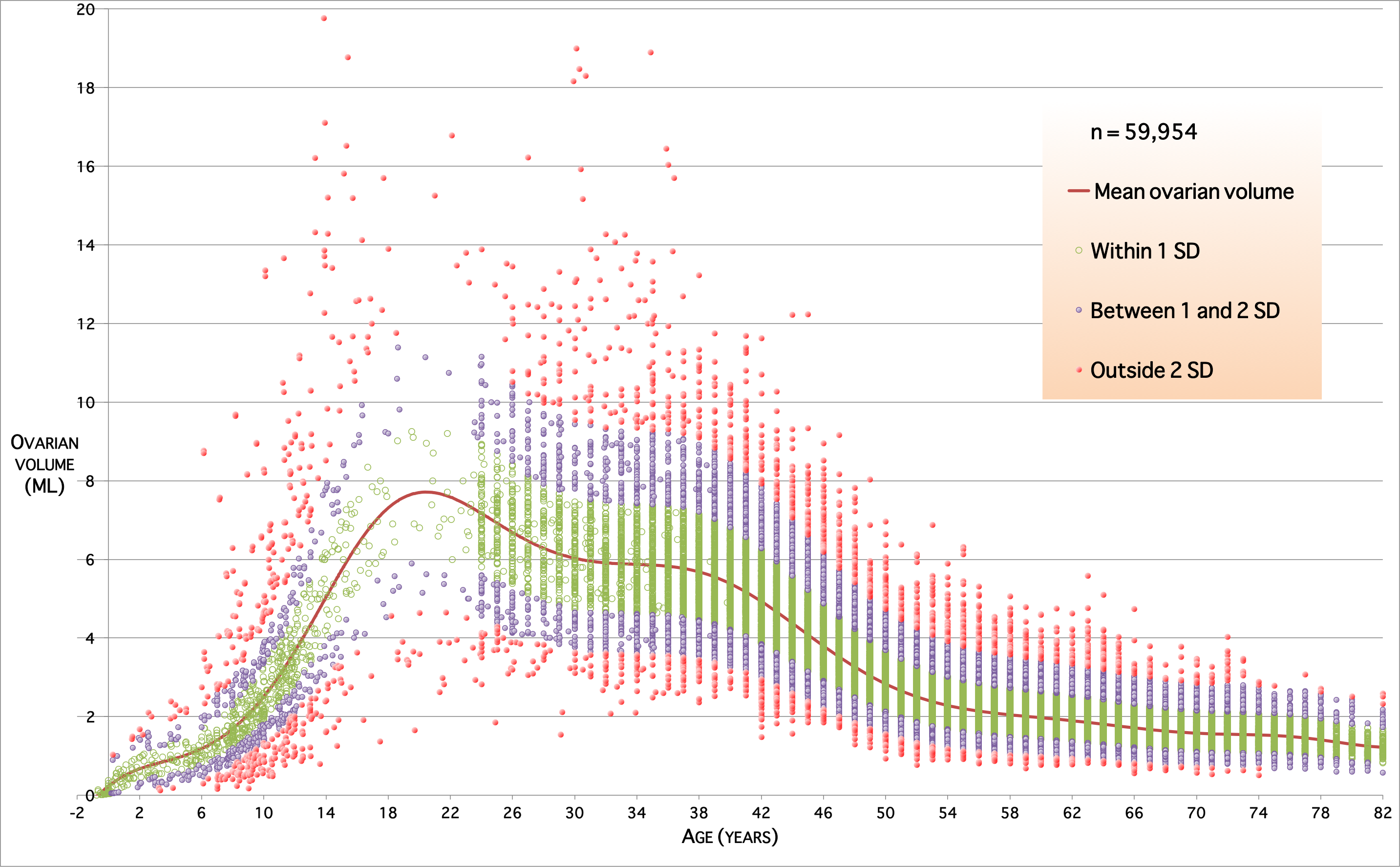}
\end{center}
\caption{
{\bf The normative validated model of ovarian volume throughout life.}  The red line is predicted mean ovarian volume in millilitres for any age. Colour bands indicate ranges within $\pm 1$ standard deviation from mean, within $\pm 1$ and $\pm 2$ standard deviations, and outside $2$ standard deviations.
}
\label{F02}
\end{figure}

The log-unadjusted predictive normative model is shown in Figure  \ref{F02}. This shows the mean volume per ovary in millilitres (mL) for the healthy human population, together with prediction intervals at $\pm 1$ and $\pm 2$ standard deviations (SD). Approximately 68\% of ovarian volumes are expected to lie within $\pm 1$ SD of the mean; approximately 95\% within $\pm 2$ SD of the mean. Mean and normative ranges for ovarian volumes are given for ages from birth to 50 years in Table \ref{Tnormative}. Our model shows that in the average case ovarian volume rises from 0.7 mL (95\% CI 0.4 -- 1.1 mL) at 2 years of age to a peak of 7.7 mL (95\% CI 6.5 -- 9.2 mL) at 20 years of age and declines throughout life to about 2.8mL (95\% CI 2.7 -- 2.9 mL) at the menopause.

 \begin{table}[!ht]
\caption{
\bf{Ovarian volumes by age}}
\begin{tabular}{c|ccccccc}
        \hline
Age	&	3SD below	&	2SD below	&	1SD below	&	Mean ovarian vol.	&	1SD above	&	2SD above	&	3SD above	\\ \hline
0	&	0.0	&	0.0	&	0.0	&	0.2	&	0.5	&	0.8	&	1.3	\\
2	&	0.0	&	0.1	&	0.4	&	0.7	&	1.0	&	1.5	&	2.1	\\
4	&	0.0	&	0.3	&	0.6	&	0.9	&	1.3	&	1.8	&	2.5	\\
6	&	0.2	&	0.5	&	0.8	&	1.2	&	1.7	&	2.3	&	3.0	\\
8	&	0.5	&	0.8	&	1.2	&	1.7	&	2.3	&	3.0	&	3.9	\\
10	&	0.9	&	1.3	&	1.9	&	2.5	&	3.3	&	4.3	&	5.4	\\
12	&	1.5	&	2.1	&	2.8	&	3.7	&	4.7	&	6.0	&	7.5	\\
14	&	2.3	&	3.0	&	3.9	&	5.0	&	6.4	&	8.0	&	10.1	\\
16	&	3.0	&	3.9	&	5.0	&	6.4	&	8.0	&	10.0	&	12.5	\\
18	&	3.5	&	4.5	&	5.8	&	7.3	&	9.2	&	11.4	&	14.2	\\
20	&	3.7	&	4.8	&	6.1	&	7.7	&	9.6	&	12.0	&	15.0	\\
22	&	3.7	&	4.7	&	6.0	&	7.6	&	9.5	&	11.9	&	14.7	\\
24	&	3.5	&	4.5	&	5.7	&	7.2	&	9.0	&	11.2	&	14.0	\\
26	&	3.2	&	4.1	&	5.3	&	6.7	&	8.4	&	10.5	&	13.1	\\
28	&	3.0	&	3.9	&	4.9	&	6.3	&	7.9	&	9.9	&	12.4	\\
30	&	2.8	&	3.7	&	4.7	&	6.0	&	7.6	&	9.5	&	11.9	\\
32	&	2.8	&	3.6	&	4.6	&	5.9	&	7.5	&	9.4	&	11.7	\\
34	&	2.7	&	3.6	&	4.6	&	5.9	&	7.4	&	9.3	&	11.6	\\
36	&	2.7	&	3.6	&	4.6	&	5.8	&	7.4	&	9.2	&	11.5	\\
38	&	2.6	&	3.5	&	4.5	&	5.7	&	7.2	&	9.0	&	11.3	\\
40	&	2.5	&	3.3	&	4.2	&	5.4	&	6.8	&	8.6	&	10.7	\\
42	&	2.2	&	3.0	&	3.8	&	4.9	&	6.3	&	7.9	&	9.9	\\
44	&	1.9	&	2.6	&	3.4	&	4.4	&	5.6	&	7.1	&	8.9	\\
46	&	1.6	&	2.2	&	2.9	&	3.8	&	4.9	&	6.2	&	7.8	\\
48	&	1.3	&	1.8	&	2.5	&	3.3	&	4.2	&	5.4	&	6.8	\\
50	&	1.1	&	1.6	&	2.1	&	2.8	&	3.7	&	4.7	&	6.0	\\ \hline
       \end{tabular} 
\begin{flushleft}
Normative values for ovarian volumes in millilitres for ages from birth through 50 years at two year stages. SD below and above refer to standard deviations below and above mean predicted volume.
\end{flushleft}
\label{Tnormative}
 \end{table} 

\begin{figure}[!ht]
\begin{center}
\includegraphics[width=\textwidth]{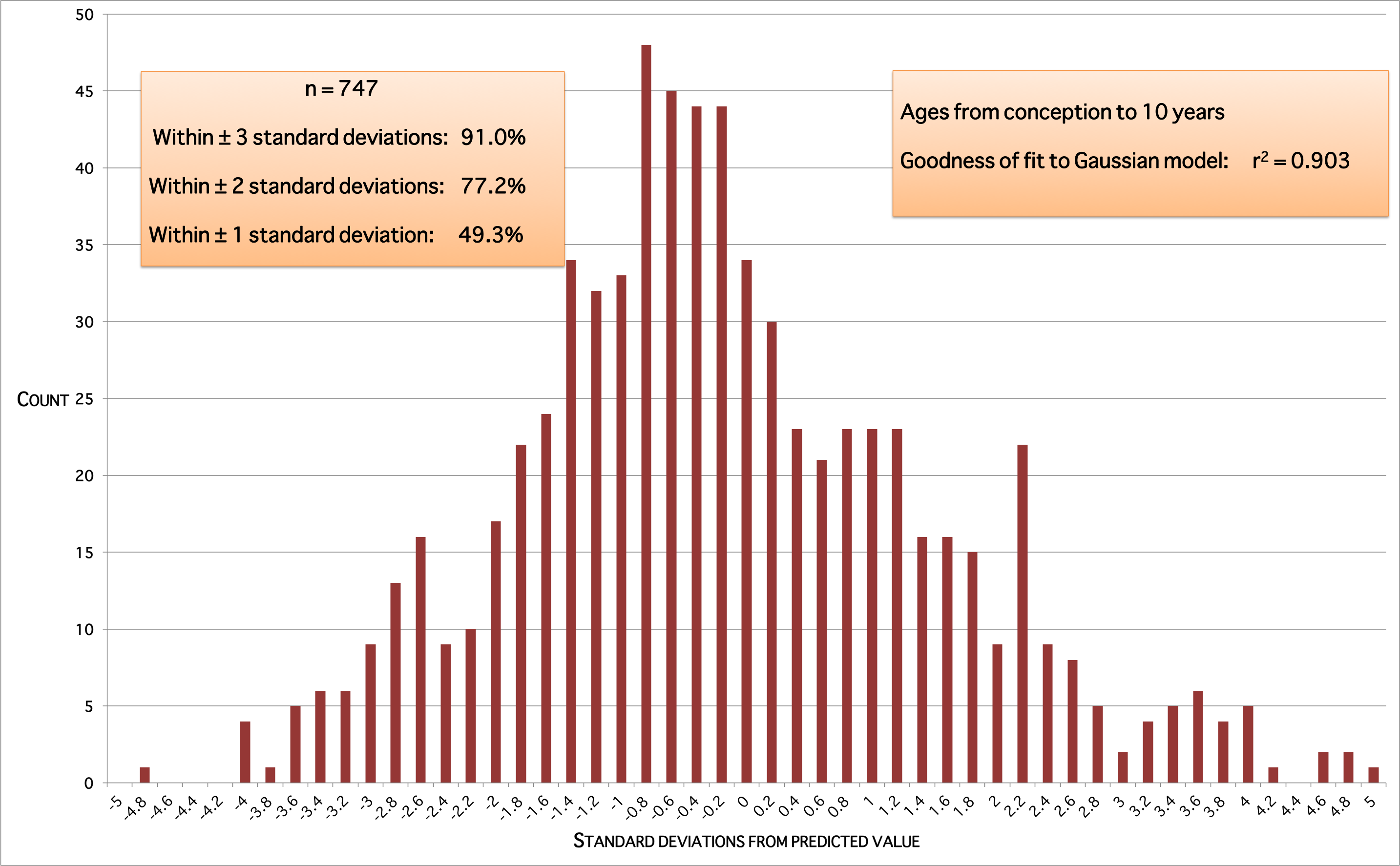}
\end{center}
\caption{
{\bf Model residuals for ages up to 10 years.}  Residuals are the squared differences between data values and predicted values for that age.
}
\label{F05}
\end{figure}

\begin{figure}[!ht]
\begin{center}
\includegraphics[width=\textwidth]{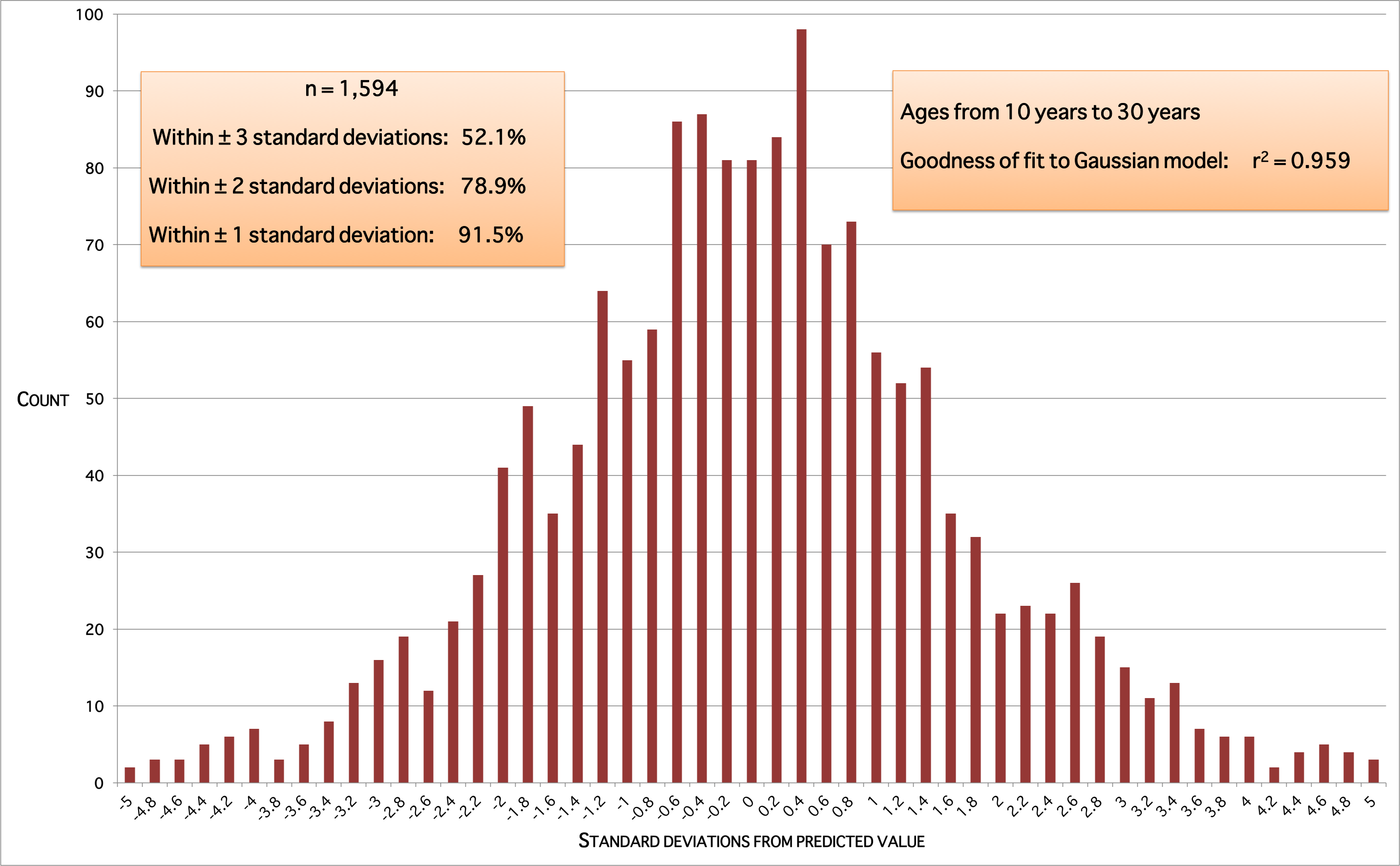}
\end{center}
\caption{
{\bf Model residuals for ages between 10 and 30 years.}  Residuals are the squared differences between data values and predicted values for that age.
}
\label{F06}
\end{figure}

\begin{figure}[!ht]
\begin{center}
\includegraphics[width=\textwidth]{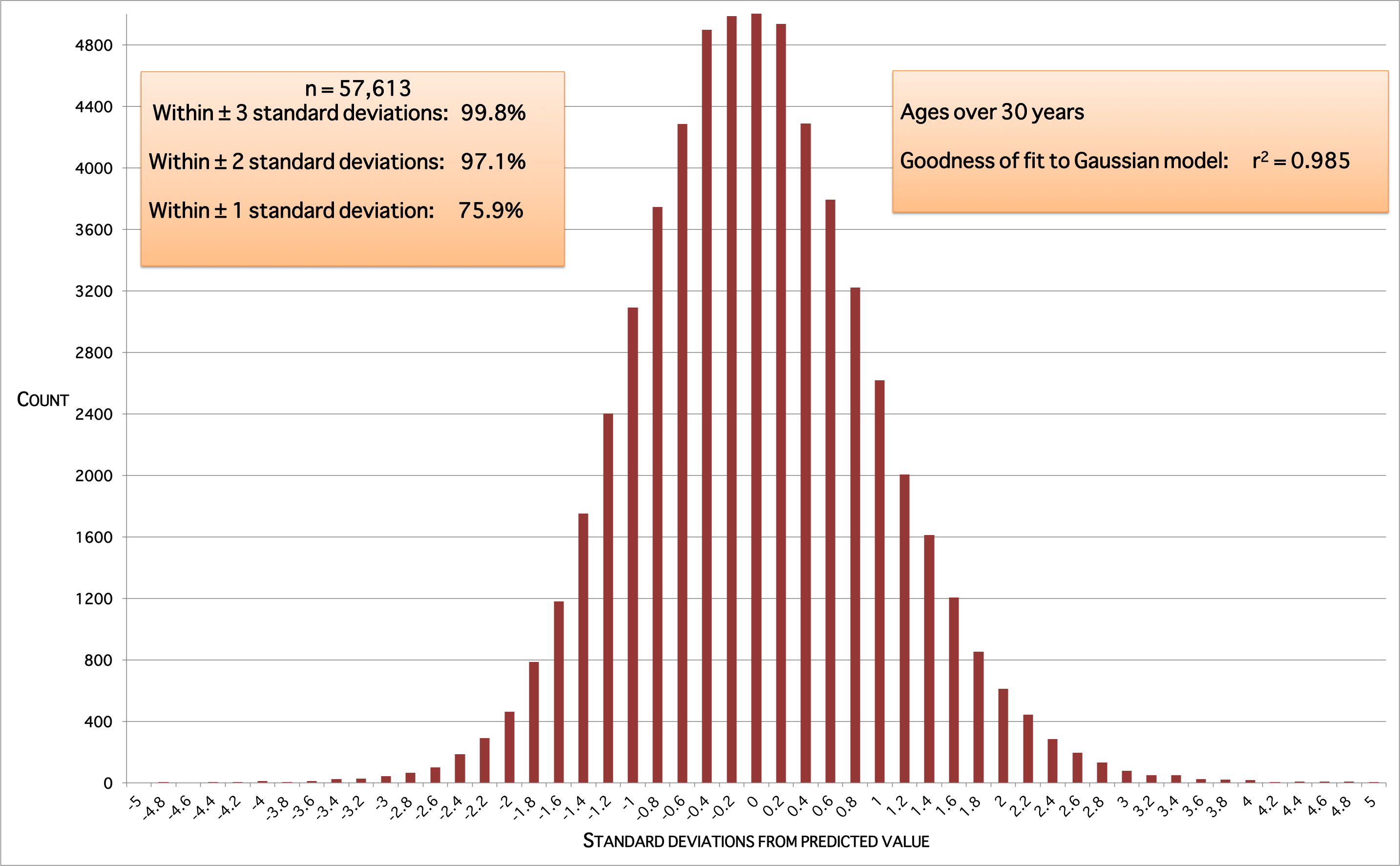}
\end{center}
\caption{
{\bf Model residuals for ages over 30 years.}  Residuals are the squared differences between data values and predicted values for that age.
}
\label{F07}
\end{figure}

The data do not support the notion of two distinct populations, PCOS and non-PCOS, giving a bimodal distribution of ovarian volumes at a given age. Model residual plots for ages up to 10 years are approximately normally distributed (Figure \ref{F05}). Model residual plots for ages 10 through 30 years (Figure \ref{F06}) and over 30 years (Figure \ref{F07}) are close to an ideal normal distribution. 

When the data is censored to remove 444 values over 10 mL, in line with the Rotterdam criteria for PCOS\cite{Rotterdam2004,Rotterdam2004a}, the model changes slightly both in qualitative and quantitative terms, with a coefficient of determination $r^2 = 0.69$ for both models. The censored-data model is the same as the full-data model for young and old ages -- average volume 0.7 mL (95\% CI 0.4 -- 1.1 mL) at 2 years and 2.8mL (95\% CI 2.7 -- 2.9 mL) at age 50 years. The censored-data model has a lower peak predicted ovarian volume in the average case, 6.4mL (95\% CI 5.4 -- 7.6  mL), with the peak occurring one year later at 21 years. This lower peak is outside the 95\% confidence interval 6.5 -- 9.2 mL for the full model peak, suggesting a statistically significant difference between the two peak values.

\section*{Discussion}

We have described and validated the first normative model that describes ovarian volume in healthy females from conception to 82 years. The model has a coefficient of determination $r^2 =0.69$ indicating that 69\% of the variation in ovarian volumes throughout life is due to age alone.  Ovarian volume rises through childhood and adolescence and is maximal in the average woman at 20 years of age, declining thereafter towards the menopause and beyond. 

Transvaginal ultrasound evaluation has been used as an indirect assessment of ovarian reserve in adult sexually active females \cite{Broekmans1998}. We have previously shown a strong positive correlation ($r=0.89$) between NGF numbers and ovarian volume from ages 25 to 51 years  \cite{Kelsey2012}, i.e. during the time that both are declining. Our normative model now adds to this by showing a steady rise in ovarian volume from birth (Figure \ref{F02}) with a modest acceleration around the onset of puberty (age 9--10 years).  The major contribution to ovarian volume before puberty is likely to be stromal growth; while small antral follicles are present in the ovaries of prepubertal girls of all ages \cite{Linternmoore1974}, larger follicles are not found while serum gonadotrophin concentrations remain low. After menarche and the onset of ovulation the major contribution to changing ovarian volume is likely to be the number and size of the antral follicles present.

Human growth in childhood is described as three additive and partly superimposed components: infancy, childhood and puberty \cite{Tse1989}. Each component appears to be controlled by distinct biological mechanisms. The infancy component is largely nutrition dependent, the childhood component is mostly dependent on growth hormone (GH) and the pubertal component depends on the synergism between sex steroids and GH. The slow rise in ovarian volume throughout mid-childhood (Figures \ref{F01} and \ref{F02}) followed by an increase in ovarian volume during the pubertal years suggests that GH, in addition to sex steroids, may have an important role in determining ovarian size (and possibly function) in the early and late childhood years. A role for GH in determining ovarian size and volume during childhood and puberty is suggested by data from Bridges {\it et al.} 1993 who studied girls with growth disorders: GH insufficiency, skeletal dysplasia, and tall stature \cite{Bridges1993}. Total ovarian volume of untreated GH-insufficient girls was significantly less than that of GH-insufficient girls on GH treatment, girls with skeletal dysplasia on GH treatment, and girls with tall stature. They also found that tall girls had significantly greater ovarian volume than either of the GH-treated groups.

The measurement of ovarian volume has been found to be useful in a wide range of disorders in children and young females. Measurement of ovarian volume is an accurate diagnostic tool for adolescent girls with irregular menses. In the majority of these girls, enlarged ovaries are associated with polycystic ovary syndrome (PCOS) \cite{Herter1996} and ovarian volume is part of the diagnostic criteria for that condition \cite{Rotterdam2004,Rotterdam2004a}. 
 
Recent results suggest that antral follicle counts have better discriminatory performance than ovarian volume \cite{Lujan2013}. We therefore censored our dataset to exclude all women with ovarian volume greater than 10mL. As the descriptions of subjects included in the original references varies, women with PCOS -- or asymptomatic women whose ovaries had polycystic ovary morphology (PCOM) may have been included. In the largest data source, PCOS was not actively excluded: `` ...Patients with a solid or cystic ovarian ovarian tumor detected by sonography were excluded from this investigation since the purpose of this study was to determine normal ovarian volume..."  \cite{Pavlik2000}. Excluding these data points resulted in a reduction in the peak average ovarian volume, as would be expected, and a slight increase in the age at which the peak was reached. Importantly, our analysis does not not shed light on the validity of the criteria for the diagnosis of PCOS.

Girls with precocious puberty have significantly increased ovarian volumes compared with a normal population \cite{Bridges1995} and ovarian volume has been proposed as a useful discriminator between central precocious puberty and premature thelarche \cite{Carel2009}. Furthermore, measurement of ovarian volume is a useful index with which to assess the efficacy of treatment of central precocious puberty with GnRH analogues \cite{Jensen1998}. 

The role of transvaginal USS as a screening test for ovarian cancer remains an important area of study \cite{vanNagell1995, Pavlik2000,vannagell2007} and transvaginal USS has an established role in the assessment and management of subfertility and in-vitro fertilization (IVF) in adult women \cite{Lass1999,Fleming2012}.  It remains difficult to assess ovarian reserve in adolescents and young women with cancer due to the considerable age-related changes in the various markers available. The measurement of ovarian volume in addition to AMH may help predict which young women are at particular risk of premature ovarian insufficiency following cancer treatment and who may therefore benefit from fertility preservation techniques \cite{Wallace2012,Brougham2012}. 

Our model is derived from data from multiple sources of the measurement of ovarian volume in otherwise healthy females. This is both a strength and a weakness of the study. The strength is that the measuring errors, both underestimating and overestimating ovarian volume, are likely to be negated as any bias is unlikely to be always in the same direction for each data source. The weakness is the heterogeneity of the values obtained from diverse sources. We cannot be certain that the measurement of ovarian volume by abdominal ultrasound, which is often difficult in young children, is as accurate as measurement by transvaginal ultrasound in older females \cite{Brett2009}. Similarly measurements taken at MRI may be different from those obtained by weighing the ovary following oophorectomy and calculating the volume from weight. 
The largest data source consists of values  imputed from a very large data source obtained by transvaginal ultrasound as part of a screening programme for ovarian cancer\cite{Pavlik2000}. 
  
This study excluded patients with a solid or cystic ovarian tumor detected by sonography, but not patients with polycystic ovary morphology. 

Our normative model of ovarian volume using data derived from multiple data sources and different methods of assessment overcomes the weakness of other studies in which only one imaging modality is used, because any potential bias in one direction is likely to be negated.

We have shown that in the average case ovarian volume rises from 0.7 mL (95\% CI 0.4 -- 1.1 mL) at 2 years of age to a peak of 7.7 (95\% CI 6.5 -- 9.2 mL) mL at 20 years of age and declines throughout later life to about 2.8 mL (95\% CI 2.7 -- 2.9 mL) at the menopause. This is the first validated normative model of ovarian volume from conception to old age; it will be of use in the diagnosis and management of a number of diverse gynaecological and reproductive conditions in females from birth to menopause and beyond. 

\section*{Materials and Methods}

The research methodology used both for data acquisition and data analysis closely follows that used to derive a validated normative model for the level of anti-M\"{u}llerian hormone (AMH) found in the blood of healthy human females for ages from conception to menopause \cite{Kelsey2011,Kelsey2012a}. 

\subsection*{Ethics statement}

Permission to perform basic science studies  on the ovarian material retrieved in Denmark was given by the Minister of Health in Denmark and by the Committee on Biomedical Research Ethics of the Capital Region on 21st September 2011 (protocol number H-2-2011-044). Written informed consent for the original human work that produced the tissue samples was obtained, and all data were anonymised prior to analysis.

\begin{table}[!ht]
\caption{
\bf{Ovarian volume data sources}}
\begin{tabular}{clc|cccc}
        \hline \multicolumn{3}{c|}{Study} &  \multicolumn{4}{c}{Statistics} \\ 
        Ref. & First author  & Year & No. ovaries & Min. age & Max. age & Median age\\ \hline 
        \cite{Kelsey2012} & Kelsey & 2012 & 58,227 & 24.0 & 85.0 &  55.0 \\ 
         \cite{Badouraki2008} & Badouraki & 2008 & 99 & 1.0 & 11.0 & 7.0 \\ 
          \cite{Razzaghy-Azar2011} & Razzaghy-Azar & 2011 & 480 & 6.1 & 13.6 & 7.0 \\ 
          \cite{Seth2002} & Seth & 2002 & 92 & 8.0 & 15.0 & 11.5 \\ 
          \cite{Holm1995} & Holm & 1995 & 165 & 5.9 & 25.4 & 13.9 \\ 
         \cite{Ziereisen2001} & Ziereisen & 2001 & 122 & 2.0 & 15.7 & 9.3 \\ 
          \cite{Griffin1995} & Griffin & 1995 & 153 & 0.0 & 14.9 & 5.8 \\ 
          \cite{Stanhope1985}& Stanhope & 1985 & 40 & 0.8 & 13.7 & 7.3\\ 
         * & Wilkinson & 2012 & 98 & 2.0 & 16.7 & 13 \\ 
         * &  Andersen & 2012 & 384 & 0.5 & 39.8 & 27.5 \\
           \cite{Sforza2004} & Sforza & 2004 & 25 & -0.5 & 0.7 & 0.0 \\ 
          \cite{Simkins1932} & Simkins & 1932 &39 & -0.7 & 14 & 0.3 \\  \hline
          \multicolumn{3}{c|}{Overall} & 59,954 & -0.7 & 85.0 & 55.0 \\ \hline

      \end{tabular} 
\begin{flushleft}
The year column refers to the year of publication; * denotes our own unpublished data. 
\end{flushleft}
\label{Tdata}
 \end{table}

\subsection*{Data acquisition}

The data for this study (Table \ref{Tdata}) come from three sources: our own measurements of ovarian volume, imputation from the large-scale study by Pavlik {\it et al.} \cite{Pavlik2000} as described in \cite{Kelsey2012}, and publications in the scientific literature. Taken as a single dataset, it approximates the healthy population in terms of ovarian volume, for ages ranging from mid-term fetal to postmenopausal.

We included data from two unpublished sources. Firstly, a detailed assessment of 300 MRI examinations in children without known endocrine, chromosomal or oncological conditions that included the pelvis, yielded 49 pairs of ovaries where both ovaries were visualized and measurable in three dimensions (median age 13 years, range 2 to 16.7 years). Ovarian volumes were calculated using the prolate ellipsoid approximation formula $a \times b \times c \times \frac{\pi}{6}$. Secondly, a further 384 ovaries (median age 27.5 years, range 0.5 to 39.8 years) were weighed before cryopreservation at the University Hospital of Copenhagen. Subjects were known to have non-ovarian cancer; subjects who had received chemotherapy were excluded. Ovarian volume was estimated using the published conversion factor for ovarian tissue density: 1.00 g/mL \cite{Rosendahl2010}.

Summary statistics were extracted from Pavlik {\it et al.} \cite{Pavlik2000} for ages 24--85 years. Repeated (10-fold) parametric bootstrapping \cite{Bootstrap} was used to simulate datapoints from the published distributions to obtain a single dataset ($n = 58, 255$) that accurately reproduces the published results.

In order to obtain data from the existing literature Ð with emphasis on volumes earlier in life than the 24 years minimum age reported in Pavlik {\it et al.} \cite{Pavlik2000} Ð studies of ovarian volume in normal, healthy girls were identified using Medline and PubMed searches using the search terms Ovary, Child, Ovarian size/ volume, Normal, Healthy and Neonatal. The references of these identified studies were then reviewed, and any other relevant research papers were extracted. Papers were included if they contained ovarian volume results for healthy, normal girls with no ovarian or endocrinological abnormalities, so as to isolate data that approximate the healthy human population. Abstracts of 37 studies were identified via this method. 

After analysis of the full papers, studies were excluded if either (i) the results consisted purely of descriptive statistics, or (ii) subjects were classified by pubertal stage rather than age. Of the remaining nine studies, seven contained data measured by trans-abdominal ultrasound and plotted in graphs \cite{Badouraki2008,Razzaghy-Azar2011,Seth2002,Holm1995,Ziereisen2001,Griffin1995,Stanhope1985} while two contained tabular data (with fetal/neonatal ovaries extracted and measured/sliced to calculate volumes) \cite{Sforza2004,Simkins1932}. The data was extracted from the graphs ($n = 1,151$) using Plot Digitizer software \cite{PlotDigitizer}, and combined with the tabular data ($n = 64$). Ovarian volumes were standardised to the prolate ellipsoid approximation formula $a \times b \times c \times \frac{\pi}{6}$ since some studies used the variation $a \times b \times c \times \frac{1}{2}$.

\subsection*{Data analysis}

 Zero volume values at conception were added to the combined dataset (Table \ref{Tdata}), in order to force models through the only known volume at any age. Since variability increases with ovarian volume, we log-adjusted the data (after adding one to each value so that zero volume on a chart represents zero ovarian volume).  We then fitted 310 mathematical models to the  training data using TableCurve-2D (Systat Software Inc., San Jose, California, USA), and ranked the results by coefficient of determination, $r^2$. Each model defines a generic type of curve and has parameters which, when instantiated gives a specific curve of that type. For each model we calculated values for the parameters that maximise the $r^2$ coefficient. The Levenberg-Marquardt non-linear curve-fitting algorithm was used throughout, with convergence to 6 significant figures after a maximum of 1,500 iterations.  For each candidate model, the mean square error and $r^2$ were calculated after removing the artificial zero values at conception. 

\begin{figure}[!ht]
\begin{center}
\includegraphics[width=\textwidth]{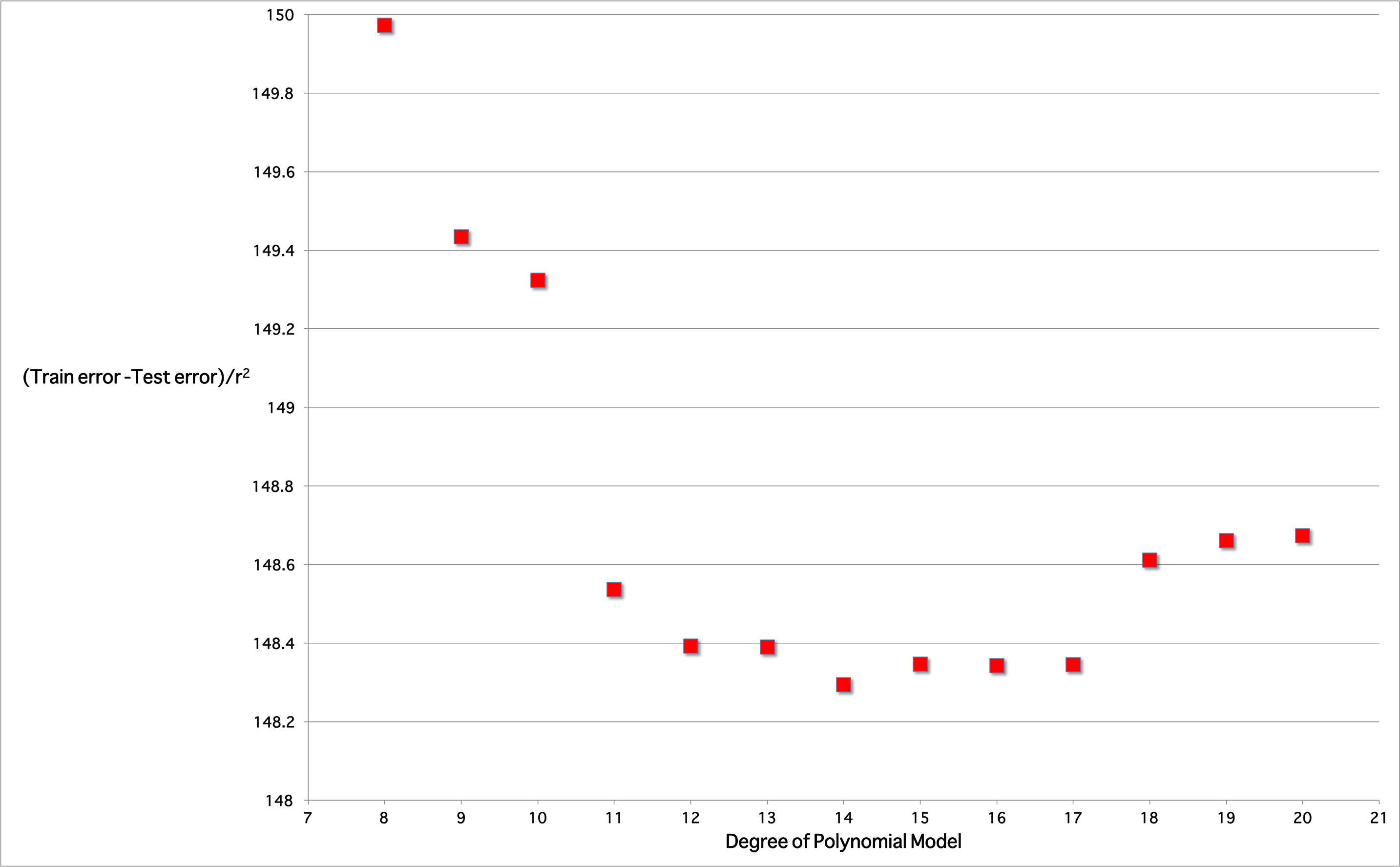}
\end{center}
\caption{
{\bf Model validation analysis.} The tradeoff between overfit and underfit for one of the five cross-validation data splits. Models with degree less than 11 are unsuitable due to low $r^2$; models with degree greater than 17 are unsuitable due to larger differences between test and training mean-squared errors. The degree 14 model is optimal. 
}
\label{F04}
\end{figure}

The best performing family of models were high precision polynomials. 5-fold cross validation was performed: the data were randomly split into 5 equally sized subsets. For each subset $S$, the other four subsets were used to train high precision polynomials of degree 8 through 20, with subset $S$ being held back as test data. 
 The mean square error of the test data was calculated and compared to the mean square error of training data for the  same model. In other words, the estimated prediction error of a model when generalized to unseen data was compared to the training error of the model. A model was considered validated if

\begin{figure}[!ht]
\begin{center}
\includegraphics[width=\textwidth]{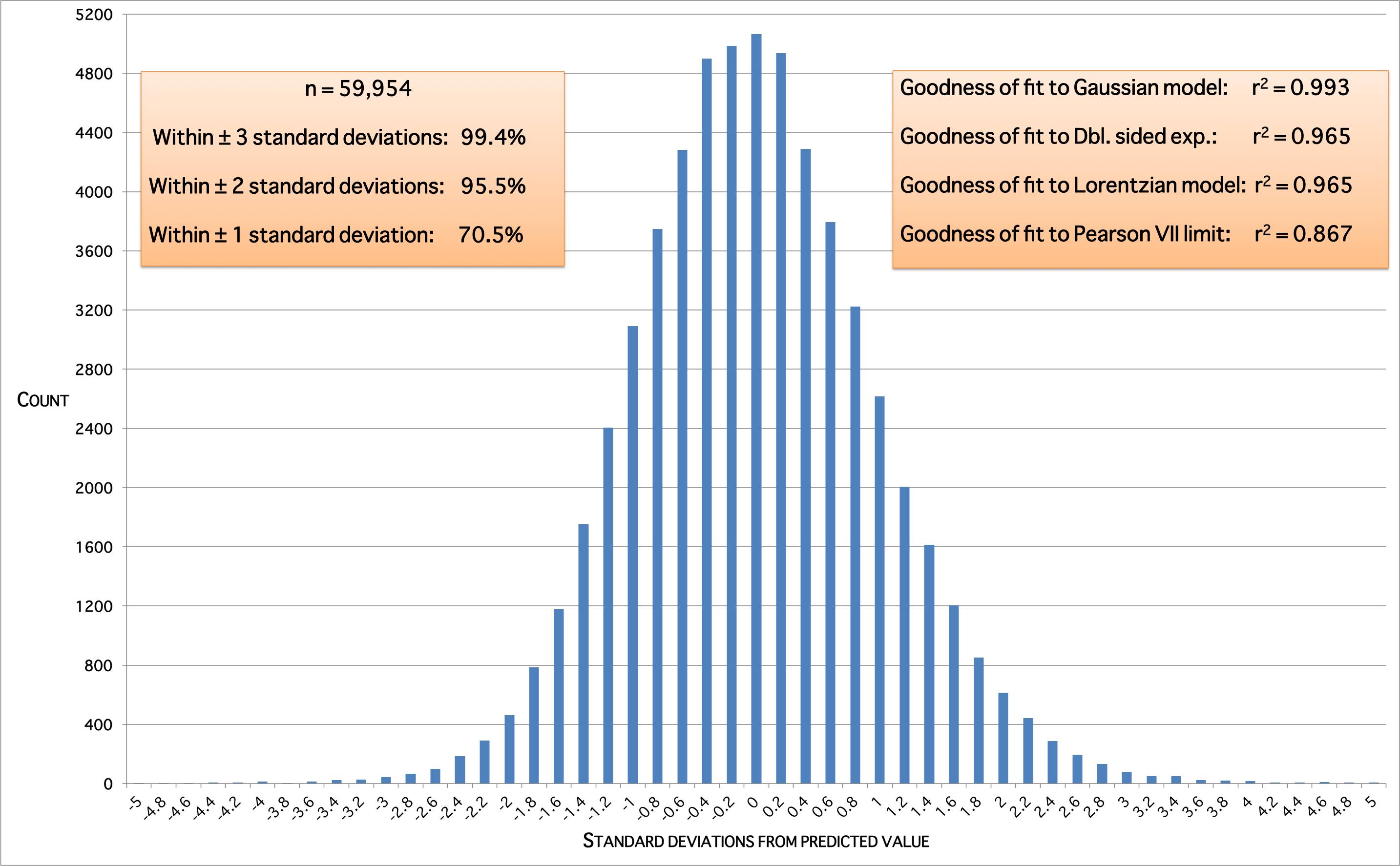}
\end{center}
\caption{
{\bf Residual distribution for the validated model.} Residuals are the squared differences between data values and predicted values for that age.  
}
\label{F03}
\end{figure}

\begin{enumerate}
\item  the residuals of the test data were approximately normally distributed (Figure \ref{F03}); and
\item   the tradeoff between high $r^2$ (denoting possible overfitting to the data) and low generalisation error (denoting possible underfitting to the data) was optimal  (Figure \ref{F04}).
\end{enumerate}

We tested for bimodal volume distributions that would suggest distinct PCOS and non-PCOS sub-populations by analysis of model residuals for age ranges up to 10 years, 10 -- 30 years, and above 30 years. Normally-distributed residuals for log-adjusted values correspond with skew-normal population volumes (i.e. a single population with PCOS and non-PCOS volumes forming a smooth continuum of values). Significant variances from normality provide evidence for a distinct PCOS sub-population.

The validated model was also assessed against the Rotterdam criteria for PCOS  \cite{Rotterdam2004,Rotterdam2004a} by censoring all values above the 10 mL discriminatory cutoff volume, re-fitting the model, and comparing peak ages and volumes.

\section*{Acknowledgments} Tom Kelsey is supported by UK EPSRC grant EP/H004092/1.

\bibliography{KDWAGAW.bib}











\end{document}